\documentstyle[psfig]{mn}
\psfull
\def\ltsima{$\; \buildrel < \over \sim \;$}
\def\lsim{\lower.5ex\hbox{\ltsima}}
\def\gtsima{$\; \buildrel > \over \sim \;$}
\def\gsim{\lower.5ex\hbox{\gtsima}}

\begin{document}

\title[Iron line in GRB afterglows]
{Iron line in the afterglow: a key to unveil Gamma--Ray Burst progenitors}

\author[Lazzati, Campana, and Ghisellini]{Davide Lazzati$^{1,2}$, 
Sergio Campana$^1$, and Gabriele Ghisellini$^1$ \\
$^1$ Osservatorio Astronomico di Brera, Via Bianchi 46, I--23807
Merate (Lc), Italy \\
$^2$ Dipartimento di Fisica, Universit\`a degli Studi di Milano,
Via Celoria 16, I--20133 Milano, Italy \\
E--mail: {\tt lazzati@merate.mi.astro.it}, {\tt campana@merate.mi.astro.it},
        {\tt gabriele@merate.mi.astro.it}
}

\date{\underline{Accepted for publication in MNRAS, 3 Feb. 1999}}

\maketitle

\begin{abstract}

The discovery of a powerful and transient iron line feature in the 
X--ray afterglow spectra of gamma--ray bursts would be a major 
breakthrough for understanding the nature of their progenitors.
Piro et al. (1999) and Yoshida et al. (1999)
report such a detection in the afterglow of GRB~970508 and 
GRB~970828, respectively.
We discuss how such a strong line could be produced in the
various scenarios proposed for the event progenitor.
We show that the observed line intensity requires a large iron mass, 
concentrated in the vicinity of the burst.
The previous explosion of a supernova, predicted in the
Supranova scenario, is the most straightforward way to account for 
such a large amount of matter.
We discuss three different physical processes that could 
account for the line: recombination, reflection and thermal emission.
Among these, reflection and thermal emission may explain
the observed line features: reflection should be important if 
the remnant is optically thick, while thermal lines can be produced 
only in a thin plasma.
The recombination process requires extremely high densities 
to efficiently reprocess the burst photons, whereas this process
could work during the X--ray afterglow. 
Future key observations for discriminating the actual radiating
process are discussed.

\end{abstract}


\section{Introduction}
Piro et al. (1999) and Yoshida et al. (1999) 
report the detection of an iron emission line
in the X--ray afterglow spectrum of GRB~970508 and GRB~970828, respectively. 
The line detected in GRB~970508 is consistent with an iron $K_\alpha$
line redshifted to the rest--frame of the candidate host galaxy 
($z=0.835$, Metzger et al. 1997), while GRB~970828 has no
measured redshift and the identification of the feature with the same 
line would imply a redshift $z \sim 0.33$.
The line fluxes (equivalent widths) are 
$F_{Fe} = (2.8\pm1.1) \times 10^{-13}$~erg~cm$^{-2}$~s$^{-1}$ 
($\rm{EW} \sim 1$~keV) and $F_{Fe} = (1.5\pm 0.8) \times 
10^{-13}$~erg~cm$^{-2}$~s$^{-1}$ ($\rm{EW} \sim 3$~keV) for GRB~970508 and 
GRB~970828, respectively.
Although the significance of these features is admittedly not extremely
compelling ($\sim 99\%$~in both cases), 
the implications that they bear are so important
to justify a study on the mechanism that would produce them.
A strong iron emission line unambiguously points towards 
the presence, in the vicinity of the burster, of a few per cent of
iron solar masses concentrated in a compact region. 
Thus the presence of such a line in the X--ray afterglow spectrum
would represent the ``Rosetta Stone'' for unveiling the burst progenitor.

Three main classes of models have been proposed for the origin of 
gamma--ray bursts (GRB):
neutron star -- neutron star (NS--NS) 
mergers (Paczy\'nski 1986; Eichler et al. 1989),
Hypernovae or failed type Ib supernovae (Woosley 1993; Paczy\'nski 1998) 
and Supranovae (Vietri \& Stella 1998).
In the NS--NS model the burst is produced during the collapse of a binary 
system composed of two neutron stars or of a neutron star and a black hole. 
In this case the explosion should take place in a clean environment, 
due to the relatively large speed (up to $\sim$1000 km s$^{-1}$) 
of such systems.
Since the time required for the binary system to coalesce and merge is
of the order of a billion year, the GRB should be outside 
the original star forming region and hence in a rarefied environment.
On the contrary, in the Hypernova scenario, the burst is due to the
evolution of a massive ($\sim 100 \, M_\odot$) star, which collapses
forming a Kerr black hole, whose rotational energy is tapped
in a few seconds, producing the burst.
Hypernovae should be located in dense molecular clouds,
probably iron rich, but there should be no Hypernova remnant.

The Supranova scenario (Vietri \& Stella 1998) assumes that,
following a supernova explosion, a fast spinning neutron star is formed
with a mass that would be supercritical in the absence of rotation.
As radiative energy losses spin it down in a time--scale of months to years,
it inevitably collapses to a Kerr black hole,
whose rotational energy can then power the GRB. A supernova remnant
(SNR) is naturally left over around the burst location.

The detection of a strong iron line redshifted to the rest frame
of the GRB progenitor poses severe problems to the
NS--NS model, which could produce lines only inside the fireball.
These hypothetical lines should be blueshifted by
the bulk Lorentz factor $\Gamma$ of the fireball (M\'esz\'aros \& Rees 1998a)
and should then be detected at frequencies $\Gamma/(1+z)$ times larger. 

X--ray line emission following GRB events has been recently 
discussed in the Hypernova scenario by Ghisellini et al. (1999) and 
Boettcher et al. (1998). 
None of these works predict, with reasonable assumptions on the burst 
surrounds, iron lines strong enough to be detectable during the
X--ray afterglow.
Moreover, line emission should last over a time--scale
of years given the width of the emitting nebula.
The production of a stronger line in the Hypernova
scenario has been mentioned by M\'esz\'aros \& Rees (1998b), who consider
recombination in a relatively dense torus, formed by the
interaction of a compact companion with the pre--Hypernova
envelope.

As we will show in this letter, the Supranova scenario can easily
account for the large amount of iron rich material needed to explain
the observed line features.

This letter is organized as follows: in section~\ref{due} we derive
model independent general constraints on the ambient material, in 
section~\ref{tre} we discuss the line emission process and in 
section~\ref{qua} we draw our conclusions.

\section{General Constraints}
\label{due}

We consider a line with a flux  comparable to a typical afterglow X--ray
flux\footnote{Here and in the
following we parametrise a quantity $Q$ as $Q=10^xQ_x$ and adopt cgs
units.}: $F_{Fe}=10^{-13}F_{Fe,-13}$ 
erg cm$^{-2}$ s$^{-1}$.
This in itself constrains both the amount of line--emitting matter and
the size of the emitting region.

Assume for that the emitting region is a homogeneous
spherical shell centered in the GRB progenitor, with radius $R$ and width
$\Delta R\le R$.
The flux of the iron line cannot exceed the absorbed ionizing fluence
$q{\cal F}$ (where ${\cal F}$ is the total GRB fluence and $q$ is
the fraction of it which is absorbed and reprocessed into the line),
divided by the light crossing time of the region, $R/c$.
This, {\it independently from the line flux variability},
gives an upper limit to the size: 
\begin{equation}
R<3\times 10^{18} \, q \, \frac{{\cal F}_{-5}}{F_{Fe,-13}} \quad \hbox{cm}
\end{equation}
Since $q$ is $\sim 0.1$ at most (Ghisellini et al. 1999), 
the emitting region is very compact,
ruling out emission from interstellar matter, even assuming 
the large densities appropriate for star forming regions.

The total line photons produced at 6.4--6.9 keV in $10^5\, t_5$
seconds, for a GRB located at $z=1$\footnote{The cosmological 
parameters will be set throughout this letter to
$H_0 = 65$~km~s$^{-1}$~Mpc$^{-1}$, $q_0=0.5$ and $\Lambda=0$.}, 
are $\sim 3\times10^{57}F_{Fe,-13}\, t_5$. 
This means that, for a reasonable amount of iron, each atom has 
to produce a large number of photons. 
For this reason, we call $k$ the number of photons produced by a 
single iron atom and we use this parameter to constrain the required mass:
\begin{equation}
M_{Fe} \sim 150\,F_{Fe,-13}\; \frac{t_5}{k} \quad M_\odot \label{miron}
\end{equation}
The parameter $k$ depends on the details of the assumed scenario, but
general limits can be set.
If we neglect thermal processes, which will be discussed in more detail 
below, any iron atom can emit photons only when illuminated by an ionizing
flux, i.e. the burst itself or the afterglow high energy tail.
Since burst light has enough power to photoionize all the matter
in the vicinity of the progenitor (see e.g. Boettcher et al. 1998),
line photons will be emitted only through the recombination process.
Thus the value of the parameter $k$ will not be larger than 
the total number of photoionizations an ion can undergo
during the burst and/or the afterglow.
For iron $K$--shell electrons, with cross section $\sigma_{K}=1.2\times
10^{-20}$~cm$^2$ we have:
\begin{equation}
k\, \lsim \, {{q\, E} \over {4\pi\, \epsilon_{ion} \, R^2}} 
\, \sigma_{K} = 
6.5\times10^6 \, {{q\, E_{52}} \over {R^2_{16}}}
\label{kappa}
\end{equation}
\noindent where $E$ is the total energy emitted by the burst and/or afterglow
and $\epsilon_{ion}$ the energy of a single ionizing photon.

Inserting equation~\ref{kappa} in equation~\ref{miron}, 
we obtain a lower limit on the iron mass $M_{Fe} \gsim 2.3 \times 10^{-5}$ $
F_{Fe,-13} \; t_5\; R_{16}^2/(q\, E_{52})\,
M_\odot$ which corresponds to a total mass:
\begin{equation}
M \, \gsim 0.013 \, 
{F_{Fe,-13} \; t_5\; R_{16}^2 \over q\,A_{\odot} E_{52}} \quad 
M_\odot
\label{minmass}
\end{equation}
i.e. a tenth of solar mass for $q \sim 0.1$ and $A_{\odot}=1$, where
$A_{\odot}$ is the iron abundance in solar units.
These general requirements about the mass and its location 
exclude that the iron line can be emitted by interstellar material,
even if made denser by a strong pre--Hypernova wind
\footnote{Assuming a wind of $\dot m_{wind}=10^{-4}$ solar masses per year
and a wind velocity $v=10^8$ cm s$^{-1}$,
the total mass within a radius $R$ is 
$M=\dot m_{wind} R/v = 3.2\times 10^{-3} 
\dot m_{wind,-4} R_{17}/v_8\,\, M_\odot$.}.

If such a large amount of mass were uniformly spread around the burst
location, it would completely stop the fireball. 
In fact (see e.g. Wijers, Rees \& M\'esz\'aros 1997) the fireball is 
slowed down to sub--relativistic speeds when the picked up mass equals 
the initial rest mass of the fireball. 
With a typical baryonic load of $\sim 10^{-(4 \div 6)} M_\odot$,
the mass predicted in equation~\ref{minmass} would stop the 
fireball after an observer time $t \sim \Gamma^{-2} R/c \sim
3 \times 10^4 \, \Gamma_1^{-2}$~s, i.e. almost one day.
Any surviving long wavelength emission should then decay exponentially 
in the absence of energy supply.
The fireball synchrotron model have been applied to GRB~970508
by Wijers \& Galama (1999) and Granot, Piran and Sari (1999). Despite
the differences of their results, both find an ambient density 
$n < 10$~cm$^{-3}$, nine orders of magnitude lower than 
the density required for the line production (see below).
The only way to reconcile a monthly lasting power--law optical afterglow 
with iron line emission is through a particular geometry, 
in which the line of sight is devoid of the remnant matter.
Therefore, the matter distribution must be anisotropic,
with the bulk of the mass located outside the line of sight of the
burst (see Fig. 1), even if the covering factor of this matter must be 
significant to reprocess a sufficient fraction of the primary burst 
photons in the line.

\section{Line emission processes}
\label{tre}
We assume that the line emitting region is located at a distance $R$ 
from the bursts, has a width $\Delta R$, density 
$n=(M/m_p)/(4\pi R^2 \Delta R)$ and scattering optical depth
$\tau_T=\sigma_T n \Delta R=(M/m_p)/(4\pi R^2)$.
These values must satisfy the constraints derived in section~\ref{due};
in addition, the first two processes discussed below require that the optical
depth is in the range 0.1--1, to let the material absorb enough
energy without smearing too much the iron line.
Consistent values are a solar mass located at $R\sim 10^{16}$ cm,
which gives $\tau_T = 0.6 (M/M_\odot)/R^2_{16}$, and
a particle density $n=9.5\times 10^9 (M/M_\odot)/(R_{16}^2\Delta R_{14})$.

The third process discussed below requires instead $\tau_T>1$,
implying $R<8\times 10^{15}(M/M_\odot)^{1/2}$ cm.

Since the line emitting material may well be a young supernova remnant, 
we allow for a high iron abundance of the plasma.

\subsection{Multiple photoionizations and recombinations in an optically
thin shell}
\label{mion}

If the plasma can remain cold and dense enough, burst photons
can be reprocessed into line photons through recombination. 
When the plasma is illuminated by burst photons,
iron atoms are fastly ionized and a line photon is produced
each time an electron recombines. 
Since burst photons rapidly re--ionize the hydrogenoid iron atom, 
the process can be very efficient and $k$ can be large.
Since the re--ionization time is very fast ($\sim 10^{-5}$~s for a typical 
burst flux), it is the recombination time that 
measures the efficiency of the line emitting process.
In this case $k = t_{ill} / t_{rec}$, where $t_{ill}$ is the illumination 
time and $t_{rec}$ is the mean recombination time.
Solving equation~\ref{miron} for the $k$ coefficient and substituting the 
iron mass $M_{Fe}$ with the total mass of the shell $M$ we obtain:
\begin{equation}
t_{max,rec} \lsim 1.3\times10^{-5}  F_{Fe,-13} A_\odot 
{{M}\over{M_\odot}} {{t_{ill}}\over{t_5}}
\label{trecmin}
\end{equation}

The recombination time of an hydrogenic ion of atomic number $Z$
in a thermal plasma is $t_{rec} = (\alpha_r \, n)^{-1}$ (Verner \& 
Ferland 1996),
where $n$ is the electron density and the recombination coefficient 
$\alpha_r$ is given by (Seaton 1959; see also Arnaud \& Rothenflug 1985; 
Verner \& Ferland 1996):
\begin{equation}
\alpha_r(Z,T) = 5.2\times 10^{-14} \; Z\, \lambda^{1/2} \, 
\left[0.429+0.5 \ln(\lambda)+{0.496\over \lambda^{1/3}}\right]
\end{equation}
\noindent where $\lambda = 1.58\times10^5\, Z^2 \, T^{-1}$.
During the burst, the Compton temperature $T_c$ of the plasma is bound 
to be large due to the high typical energies of burst photons and to 
the relative inefficiency of radiative cooling processes. 
The free--free cooling time is of the order of $10^5$~s. 
For a typical burst spectrum we have $T_c \sim 10^8$~K. 
The recombination time turns out to be  $t_{rec} \sim 10^{2} \, n_{10}^{-1}$~s, 
while equation~\ref{trecmin}, with an illumination time of 100~s,
gives $t_{max,rec} \lsim 10^{-2}$~s. 
We hence conclude that recombination cannot be effective during the burst.
During the afterglow, Inverse Compton losses cool the plasma
efficiently, leading to a lower Compton temperature.
For GRB~970508, the observed optical/X--ray spectrum half a day after
the burst gives $T_c \sim 6 \times 10^6$~K.
This yields a shorter recombination time, $t_{rec} \sim 10 \, n_{10}^{-1}$~s, 
to be compared with the value $t_{max,rec} \lsim 1$~s, 
obtained from equation~\ref{trecmin} for a shell of unit solar mass and 
solar iron abundance.
We conclude that, during the afterglow, a shell with several
solar masses and/or high iron abundance could produce the observed line 
through the recombination process.

\begin{figure*}
\hspace{2cm}
\psfig{figure=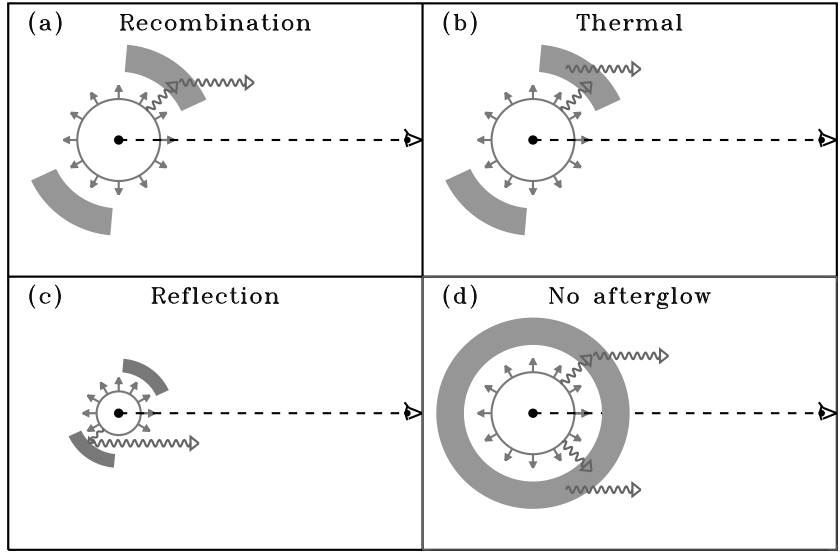,width=11cm}
\caption{{Cartoon illustrating the different emission mechanisms discussed
in this paper. Shaded regions represent the supernova remnant, circles
(with arrows) represent the fireball. In a) photons from the bursts
photoionize iron atoms in the remnant shell, which recombine many times
during the burst. In b) the shell has been heated by the burst photons,
and emits thermally. In c) the shell, more compact than in the other cases,
produces an iron fluorescent line, and a Compton reflection continuum.
The last panel d) shows that if the remnant has a covering factor of unity,
there should be no standard afterglow emission, since the fireball is
suddenly stopped by the remnant.}
\label{figuno}}
\end{figure*}

\subsection{Thermal emission from the surrounding shell}

This process should become efficient after the burst has passed,
leaving behind a thermal plasma with a temperature $T \sim 10^8\,T_8$~K.
This plasma is in the same conditions of the intra cluster medium (ICM)
in cluster of galaxies, systems that emit a strong $6.7$~keV iron line
(Raymond \& Smith 1977; Sarazin 1988).
A key question to solve if we want to apply ICM computations in this 
case is whether the collisional ionization equilibrium holds in our
plasma. In the very first time, soon after the burst photons
have passed, the iron will be almost completely ionized, and a recombination
time $t_{rec}$ is needed to reach equilibrium. From section~\ref{mion}
we have that $t_{rec} \sim 100$~s for standard shell parameters.
Since this time is very short compared to the equilibrium 
cooling time of the plasma ($t_{cool} \simeq  2.3\times10^5\,
n^{-1}_{10}\, T^{1/2}_8$~s), we can assume collisional equilibrium 
to compute the iron line intensity.

The equivalent width of the line in a solar abundance plasma
has been carefully computed by Bahcall \& Sarazin (1978) (see in particular 
their Figure 1) and ranges from several tens of eV at high ($5\times 10^8$~K)
temperatures to $\sim 2$~keV at $2.5 \times 10^7$~K. 
A very weak line is expected for temperature lower than $5\times10^6$~K.
For temperatures larger than $5\times10^7$~K the EW dependence on 
temperature can be reasonably approximated as a power law.
Assuming an iron abundance 10 times solar we have:
\begin{equation}
\hbox{EW}(T) \simeq 3.8 \, T_8^{-1.9}\quad {\rm keV}\quad (T_8 \geq 0.5)
\label{ewt}
\end{equation}
\noindent Taking into account the spectral energy density of the 
bremsstrahlung continuum at 6.7~keV, we obtain a line luminosity of:
\begin{equation}
L_{Fe} \simeq 8 \times 10^{44} \, \exp\left(-{0.8\over T_8}\right)
\, \left({M \over M_\odot}\right)^2 \, 
V_{47}^{-1} \, T_8^{-2.4} \; \hbox {erg s$^{-1}$}
\label{llin}
\end{equation}
\noindent for a shell of volume $V=10^{47} \, V_{47}$~cm$^3$.
For a $z=1$ burst we obtain a flux:
\begin{equation}
F_{Fe} \simeq 2.5 \times 10^{-14} \, \exp\left(-{0.8\over T_8}\right)
\, \left({M \over M_\odot}\right)^2  \, V_{47}^{-1} \, T_8^{-2.4} \;
%\hbox {erg cm $^{-2}$ s$^{-1}$}
{{\hbox{erg}}\over{\hbox{cm$^2$ s}}}
\label{flin}
\end{equation}
Therefore a shell of several solar masses, typical
for many type II SN (see Raymond 1984; Weiler \& Sramek 1988; Woosley 1988; 
McCray 1993), at a temperature slightly below $10^8$ K can produce a line 
flux of $10^{-13}$ erg cm$^{-2}$ s$^{-1}$ for $z=1$ bursts.
The EW with respect to the underlying bremsstrahlung radiation would be a few
keV, but any other emission component (e.g. afterglow emission) 
would decrease the line EW.
Note that the predicted X--ray bremsstrahlung continuum has a flux 
$F_{ff} \sim 6 \times 10^{-14} \, (M/M_\odot)^2 \, T_8^{1/2} \, 
V_{47}^{-1}$~erg~cm$^{-2}$~s$^{-1}$,
a value comparable with a typical burst afterglow X--ray flux, especially
if $M\sim$~a few $M_\odot$.
The line emission process can be stopped after about one day,
if the afterglow photons enhance the
plasma cooling via inverse Compton, lowering the temperature down to
less than $10^7$~K. Line emission can also be quenched by the re--heating 
produced by the incoming fireball.

\subsection{Reflection}

In Seyfert galaxies we see a fluorescence 6.4 keV iron line produced
by the relatively cold ($T<10^6$ K) accretion disk, illuminated
by a hot corona, which provides the ionizing photons 
(e.g. Ross \& Fabian 1993).
The EW, if the observer receives both the hot corona emission and
the line photons, is of the order of 200 eV if the disk intercepts 
$\sim 1/2$ of the hard X--rays.
In such systems the radiation energy density 
$U_r\sim 10^8L_{45}/R_{13}^2$~erg~cm$^{-3}$, 
similar to the radiation energy density at $R=10^{15}$ cm
from the burst.
It is therefore conceivable that a similar mechanism can work also for GRB,
if there exists a dense material in the vicinity of the burst
(M\'esz\'aros \& Rees 1998b).
In the case of GRB, the equivalent width could be much larger,
since the reflected component (line and Compton bump) is observed
when the burst has faded and only the much weaker afterglow
contributes to the continuum.
In this case, besides a scattering optical depth $\tau_T>1$, we require
a size large enough to allow the line being emitted even $\sim$~one day
after the GRB event (i.e. $R \gsim 10^{15}$ cm).

In this model the emission line is produced only during the burst event,
but in the observer frame it lasts for a time $R/c$.
The observed luminosity of the Compton reflection
component is equal to the $\sim$~10\% 
of the absorbed energy, divided by the time $R/c$: 
$L\sim 3\times 10^{45} \,E_{abs, 51}/R_{15}$
erg s$^{-1}$.
The luminosity in the iron line (see e.g.~Matt, Perola \& Piro 1991) 
is roughly 1\% of this, times the iron abundance in solar units.
Therefore the Compton reflection component 
can contribute to the hard X--ray afterglow emission 
and the iron line can have a luminosity up to 
$3\times 10^{44} \, A_\odot \, E_{abs, 51}/R_{15}$~erg s$^{-1}$, 
corresponding to fluxes up to $10^{-13}$ erg cm$^{-2}$ s$^{-1}$ 
for a $z=1$ burst.

\section{Discussion}
\label{qua}

We have discussed three possible mechanisms for the production of
a strong iron line, visible during the X--ray afterglow emission of GRBs.
All mechanisms require the presence of a large amount of iron in a compact
region. Both the general constraints derived in section~\ref{due}
and the limits due to the particular emission processes discussed
in section~\ref{tre} point towards the presence of more than a 
solar mass of matter, iron rich, in close vicinity with the burst
location. The more natural astronomical scenario in which such
conditions are found is the young remnant of a supernova, exploded
several months before the burst onset. In fact, with a radial velocity 
of the ejecta $v_{ej} = 10000$~km s$^{-1}$, a monthly lived SNR has a radius
of $R \sim 2.6 \times 10^{15}$~cm. A young SNR surrounding the burst
is predicted by the Supranova scenario (Vietri \& Stella 1998).

The other strong general requirement concerns the special geometry
needed if we want to explain the presence, in the same burst, 
of both a strong iron line and an optical afterglow (if interpreted
as due to a decelerating fireball).
Since the line emitting plasma receives the burst radiation from a 
different orientation than our line of sight, the iron emission line 
is a powerful tool to measure how isotropic the burst emission is.

We find that the multiple ionization and recombination scenario
has difficulties in reconciling the low temperature required to have
a fast recombination with the large heating due to the burst flux.
However, during the afterglow, the longer illumination time and
the lowest plasma Compton temperature allow a stronger emission line 
produced by recombination, marginally consistent with the Piro et al. (1999)
and Yoshida et al. (1999) observations.
The two other alternatives (i.e. thermal emission and reflection) 
are more promising and not mutually exclusive.
The prevalence of one over the other mechanism depends on the set up
of the system: compact regions, possibly corresponding to
very young supernova remnants ($\sim 1$~month), would produce iron line
photons by fluorescent reflection, while somewhat more extended regions,
corresponding to less young remnants ($\sim 1$~year), could produce
thermal emission.
Much weaker line fluxes, but lasting for a longer time, can be produced
by more relaxed systems (i.e. supernovae exploded more than one
year before the burst).

If the emission line is produced by a thermal plasma, its duration is of 
the order of the cooling time, since this is likely to be longer than the 
light crossing time $R/c$.
On the other hand, as discussed above, the iron line emission is quite
sensitive to the temperature, and can then be quenched if the emitting
material is suddenly re--heated by the incoming fireball or cooled 
by the afterglow photons.
In the first case, the line flux can decrease rapidly, to increase again
later on, once the shell has cooled again to the appropriate temperature.
This mechanism would allow relatively short lived
($\sim R/c$) lines even in the thermal emission scenario.

With the available information, it is hard to tell which is the case 
for GRB~970508 and GRB~970828. 
The first had a $~1$ keV equivalent width line whose flux apparently 
disappeared after half a day, in concomitance with the 
``rebursting" phase in the X--ray and optical bands.
The second burst had a $\sim3$~keV equivalent width line whose flux, instead, 
appeared in concomitance with a small ``rebursting'' phase.
The continuum spectra should be the sum of the power law afterglow emission 
and a bremsstrahlung spectrum (in the case of thermal emission) or a
harder (in the 10--100 keV band) Compton reflection spectrum 
(in the case of reflection).
The short duration of both lines, if real, corresponds to a 
size $R \lsim 10^{15}$~cm, implying a Thomson thick remnant
and favoring the reflection model.

As it is often the case, to be more conclusive
we must await better spectra of other bursts:
a key signature for thermal emission would be the detection of
a strong iron $K_\beta$ blend. 
In fact, this line cannot be produced by fluorescence given the 
lower photoelectric yield and is very weak even in a recombination scenario. 
At lower energies (1--3 keV, rest frame), $L$--shell iron lines, 
Mg, Si and S lines should also be visible (see Sarazin 1988). 
In the reflection scenario, afterglow spectra should show the typical 
hardening of the spectrum above a few keV, and line duration
should be short.

The possible association of GRB with supernovae has
been investigated recently in detail by Bloom et al. (1998), 
Kippen et al. (1998) and Wang \& Wheeler (1998),
following the explosion of GRB~980425, likely associated with the type Ic
SN 1998bw. Among these works, only Wang \& Wheeler (1998) find evidence for a 
connection while the other two limit to a few percent the bursts
possibly associated with supernovae.
In the Supranova scenario, however, the association of 
supernovae with bursts should suffer a time delay, variable between
few days to some years, which would smear the time correlation
between the two phenomena. 

Should the iron line possibly detected in GRB~970508 and GRB~970828
be real and confirmed by other cases, then we have a strong case
for the connection between supernovae and gamma--ray bursts.
The next generation of experiments and satellites, 
such as XMM, AXAF and ASTRO--E, will provide us with the necessary
information to draw more accurate conclusion on the puzzling
problem of the gamma--ray burst progenitor.

\section*{Acknowledgments}
We thank L. Piro and A. Yoshida for providing us a copy of their manuscripts
in advance to publication. We thank M. Vietri, S. Covino and E. Ripamonti 
for fruitful discussions during the conception and preparation 
of this work. D.L. thanks the Cariplo Foundation for financial support.

\end{document}